\documentclass[12pt]{article}
\usepackage[centertags]{amsmath}
\usepackage{footnote}
\usepackage[flushleft]{threeparttable}
\usepackage{amsfonts}
\usepackage{amssymb}
\usepackage{amsthm}
\usepackage{amsmath}
\usepackage{newlfont}
\usepackage{graphicx}
\usepackage{subfig}
\usepackage{placeins}
\usepackage{caption}
\usepackage{booktabs}
\usepackage{amsthm}
\usepackage{hyperref}
\usepackage{epsfig}
\usepackage{color}
\usepackage{latexsym}
\usepackage{graphicx}
\usepackage{booktabs}
\usepackage{array}
\usepackage{rotating}
\usepackage{xcolor}
\usepackage[table]{xcolor}
\usepackage{subcaption}
\newfont{\bb}{msbm10 at 14pt}

\usepackage{setspace}

\usepackage[left=2cm,right=2cm,top=2.3cm,bottom=2.3cm]{geometry}
\usepackage[font=small,skip=0.1pt]{caption}
\usepackage[T1]{fontenc}
\usepackage[utf8]{inputenc}
\usepackage{authblk}
\usepackage{booktabs}
\usepackage{siunitx}
\usepackage{multirow}
\date{}

\begin{document}
	\begin{center}{\Large\bf{ Analysis of Fully-Heavy and Hidden-Heavy Tetraquarks in Anisotropic Plasma Using the Generalized Fractional Derivatives  }}
	\end{center}
	
	\begin{center}
		H. M. Fath-Allah $^{1}$, M. Abu-Shady $^{2}$ and E. M. Khokha $^{3,*}$
		
		\bigskip
		$^{1}$ Higher Institute of Engineering and Technology, Menoufia, Egypt (hebamosad8@gmail.com)
		
		$^{2}$	Department of Mathematics and Computer Sciences, Faculty of Science, Menoufia University, Shebin El-kom,
		Egypt ( dr.abushady@gmail.com)
		
		$^{3}$	Faculty of Computer Science and Engineering, King Salman International University (KSIU), South Sinai, Egypt (Corresponding author: emad.khokha@ksiu.edu.eg)

	\end{center}
	\begin{abstract}
Investigating the spectroscopy and thermal stability of exotic multiquark states provides crucial insights into the fundamental confinement mechanisms of quantum chromodynamics (QCD) under extreme regimes.
This work presents an innovative study of the spectroscopy and dissociation of fully-heavy tetraquarks in an anisotropic plasma via the generalized fractional derivatives (GFD) structure. The radial Schrödinger equation (SE) with an extended Cornell potential that includes Debye screening and plasma anisotropy is solved by applying the parametric generalized fractional Nikiforov-Uvarov (PGFNU) technique. The analysis of systems consisting of $cc\bar c\bar c$ and $bb\bar{b}\bar{b}$ shows that the binding potential of the systems and the dissociation energy increase with the anisotropy parameter and decrease with temperature. The fractional  paradigm also shows that when the fractional orders are lower, the systems will be more strongly bound than in classical quantum mechanics. Furthermore, the mass spectra for all considered tetraquark configurations, including the $cc\bar{c}\bar{c}$, $bb\bar{b}\bar{b}$, $c\bar{q}c\bar{q}$, and $b\bar{q}b\bar{q}$ systems, are calculated for both ground and excited states. These calculations were performed across both the fractional and classical regimes. The findings indicate that the GFD framework provides an effective mathematical basis for modelling hadronic systems in complex contexts, and they demonstrate significant agreement with earlier theoretical studies.
	\end{abstract}
	\section{Introduction}
Tetraquarks are an interesting family of exotic hadrons that are not limited to quarks (which include quark-antiquark mesons and quark-quark-quark three-quark baryons). The observation of a significant number of charmonium-like resonances, such as: \textit{X}(3872) \cite{r1}, $Z_c$(3900) \cite{r2} and $Z_c$(4430) \cite{r3} which challenge traditional concepts of hadron spectroscopy have generated a lot of interest in these four-quark states $\textit{QQ} \bar Q \bar Q$ or $\textit{QQ} \bar q \bar q$. This recent finding \cite{r4} of the possible totally charmed tetraquarks at the di-$J/ \psi$ threshold at the Large Hadron Collider has increased the efforts of theorists to clarify the assembly, binding mechanisms and dynamics of these multiquark complexes.The diquark-antidiquark model has proven to be a valuable model in which to discuss tetraquarks. The image represents the four-quark system in a confined state of strongly correlated pairs of quarks (diquarks) and their antiparticle counterparts. This makes the four-body problem complex, since the transformation of the problem into a two-body interaction is simplified. The antitriplet diquark channel contains strong color-attractive forces, and spin-dependent hyperfine interactions which favour certain diquark configurations, leading to a hierarchical structure. Various theoretical applications have used the diquark-antidiquark model in several theoretical contexts to produce accurate predictions of the masses and binding energies of the tetraquarks as well as the decay properties of tetraquarks.
The techniques of the QCD sum rule methods relate the low-energy hadronic measurements directly to the parameters of QCD. The method was used by Agaev et al. \cite{r5}. They used the method to study fully heavy tetraquarks with axial-vector potential interaction, i.e., $QQ\bar Q \bar Q$. They did this by including the contribution of nonperturbative gluon condensate contributions to correlation functions to study both charming and bottom configurations. The method is based on the QCD dynamics to make predictions but must be careful with regard to operator product expansion and continuum contributions. 

Non-relativistic potential models are the most common way of performing tetraquark spectroscopy. These techniques typically use Cornell-type potentials that are short-range Coulomb-like interactions augmented by long-range linear confinement and spin-dependent corrections. The method applied by Ghalenovi \cite{r6} to estimate the masses of charmed and bottom tetraquarks, is that of resolving the radial SE, with confining potentials and hyperfine interaction. Extensive spectroscopic studies on totally charmed systems were carried out by Debastiani and Navarra \cite{r7} and spin-spin interaction, spin-orbit interaction and tensor forces were studied. Mistry and Majethiya \cite{r8} made systematic predictions of the masses of fully-heavy tetraquarks of the form of $QQ\bar Q \bar Q$ in both the ground and excited states. The forecasts covered scalar and vector orientations that are of relevance to the recent LHC observations. Cornell potentials were used by Lundhammar and Ohlsson \cite{r9} to analyze the meson data and to predict the masses of 24 ground-state tetraquarks, including both heavy-light and all heavy configurations.

The ground-state masses of tetraquark states of $c \bar q c \bar q$ tetraquarks in the range of 3.8-4.7 GeV correspond to the reported charmonium-like states $Z_c$ (3900), $Z_c$ (4430), and $Z_c$ (4660) \cite {r10}. Bhavsar et al. \cite{r11} used linear confinement and spin-dependent interactions to calculate the masses of concealed charm tetraquarks $c c \bar q c c \bar q, c c bar c c \bar s, c c \bar c c \bar q$. They established those states that were above the barrier of the $D$-$D^*$ and studied how easily such states could break. Electromagnetic attributes were also anticipated by the constituent quark model as presented in \cite{r12}. It calculated magnetic moments, radiative decay widths, and binding energies of systems with charm, bottom and mixed flavors.
More sophisticated dynamical theories have also given important information on the stability of tetraquarks. Mutuk \cite{r13} used the dynamical diquark model and Born-Oppenheimer approximation to study mixed-flavor tetraquarks $cc\bar b\bar b, bc\bar c\bar c, bc\bar b\bar b$. It was found that, $cc\bar b\bar b$, is located below the two-meson threshold, which means that this state is a true bound state, but that $bc\bar c\bar c$ and $bc\bar b\bar b$ are above the dissociation limit. Meng et al. \cite{r14} have used the Gaussian expansion and complex-scaling methods and the one-gluon exchange and the linear confinement potentials to solve the four-body SE. They did this by identifying confined and resonant states in different configurations of flavors and spin-parity.
Relativistic theories are useful when examining the effects that are kinematic, and it is essential when considering systems that involve light quarks, or consistency with the QCD symmetry. Ghasempour et al. \cite{r15} applied the relativistic Bethe-Salpeter equation to the Hellmann potential and spin dependent terms. Ansatz method was used to calculate the mass of tetraquarks of hidden-charm and hidden-bottom states in a systematic manner. A relativistic diquarkonium model using Cornell-type potentials and relativistic spin-dependent interactions was used by Zolfagharpour and Aslanzadeh \cite{r16} to calculate the  masses of the ground and excited state of both the scalar and axial-vector diquark configurations. Shiri et al. \cite{r17} developed a method of analysis which makes use of the Bethe-Salpeter equation with Cornell potential, which simplifies the resolution of the four-body problem by means of solutions through the ansatz method.

To compute the masses of fully heavy tetraquarks, Faustov et al. \cite{r18} used the relativistic quasipotential method, using form factors which included the internal structure of diquarks, to calculate the masses of fully heavy tetraquarks. These states were found above the two-meson thresholds, at which they mainly decays into quarkonia. The large resonance at the di-$J/ \psi$ threshold could be due to the presence of a large resonance at the $2^{++}\, c c \bar{c} \bar{c}$ ground state. Hadizadeh and Khaledi-Nasab \cite{r19} took the Lippmann-Schwinger integral equation, with regularized potentials, that combined the linear confinement and Coulomb interactions. The results showed that the masses of the ground and excited states were consistent with experimental findings. Complementary phenomenological approaches offer constraints that are not based on any single paradigm. Patel, Oudichhya and Rai \cite{r20}-\cite{r21} used Regge phenomenology to investigate mass spectra in both charmed and bottomed tetraquark forms by building up quasi-linear Regge trajectories and finding constraints on mass inequality. Anwar and Burns \cite{r22} have found the correlations of linear masses of s-wave tetraquarks, which are influenced by hidden color, spin and spatial degrees of freedom. This connected quark and diquark models gave us model-independent tools to restrict the configuration of tetraquarks. A new relativistic study \cite{r23} of the mass structure of heavy tetraquarks, with a Bethe-Salpeter framework that includes Hellmann potential, and spin-dependent coupling factor. This study also confirmed the diquark-antidiquark composite model.
Along with the vacuum qualities, the understanding of the tetraquark dynamics in the hot and dense QCD matter obtained by the heavy-ion collisions has become a crucial area of study. The high temperatures achieved in facilities such as the LHC and RHIC, where the temperatures exceed the critical barrier to deconfinement, provide unique possibilities to study the mechanisms of exotic hadron formation and dissociation. Strickland and Bazow \cite{r24} analyzed bottomonium suppression in ultrarelativistic heavy-ion collisions using a pNRQCD-based potential approach, and comparing the suppression patterns in terms of centrality, rapidity, and transverse momentum at RHIC and LHC energies, and constraining the properties of transport with experimental data on STAR and CMS. In an anisotropic hot QCD medium, Thakur et al. \cite{r25} investigated the properties of quarkonium by modifying the Cornell potential with hard-loop resumed dielectric functions. They discovered that anisotropy in momentum space will result in angular anisotropy in the potential, such that the quarkonium states are tighter bound than they would be in an isotropic medium. Pairs of quarks aligned with the anisotropy direction are seen to be bound more strongly than pairs oriented perpendicular to the anisotropy. With the solution of the four-body SE in the systems of the four-quark \cite{r26}, the system of the four-quark was examined: the behavior of fully-heavy tetraquarks in the vacuum, as well as in the highly-interacting matter at finite temperatures. They did this by applying Wigner function formalism and coalescence techniques to combine temperature dependent mass evolution and dissociation temperatures. During heavy-ion collisions at the LHC, it was anticipated that the yield of the production of the $c\bar c$ and the distribution of their transverse momentum would be higher than that measured in nucleon-nucleon collisions. Using thermal QCD sum rules, Aydin et al. \cite{r27} investigated the temperature dependent behavior of scalar fully-heavy tetraquarks, finding that the mass and decay constant of fully-heavy tetraquarks significantly dropped approaching the critical temperature, and that the thermal effects of fully-heavy tetraquarks were significantly greater than the corresponding bottom tetraquarks. Silva et al. \cite{r28} used the SE with a temperature-dependent Cornell-like potential to study the thermal stability of fully-heavy tetraquarks in the diquark-antidiquark framework, and found that they dissociate above twice the critical temperature and are not produced in a hot QCD medium.

The recent application \cite{r29} of GFD methodologies in high-energy physics has evolved into a more comprehensive framework of elucidating complex QCD phenomena in particular in studying heavy quarkonia, exotic multiquark states, and their dynamics at extreme conditions. This framework was developed by Abu-Shady and Fath-Allah via the generalized fractional analytical iteration method to determine the masses of single, double, and triple heavy baryons in a hyper-central quark model, that takes into account Coulombic, linear confining, and harmonic potentials, with and without hyperfine terms. Abu-Shady et al. \cite{r30} have used a GFD extension of the NU method to derive analytic-exact fractional energy spectra and wave functions of fully-heavy tetraquarks, and applied this method to the fractional SE, with Cornell, harmonic, and spin-dependent interactions. Their results showed that changing the fractional order can lead to drastic changes on the expected tetraquark masses in comparison to traditional quantum-mechanical systems. The framework was later expanded by Abu-Shady and Ezz-Alarab \cite{r31} to include thermodynamic analyses. They used the GFD and accurate iteration of the SE of quarkonia in quark-gluon plasma with the screening potentials. They established the energy eigenvalues, mass spectrum and dissociation temperatures of charmonium and bottomonium states.Abu-Shady \cite{r32} took a step forward by solving the radial SE in anisotropic quark-gluon plasma media using a generalized NU method to include the screening potentials. This yielded the eigenvalues of energy and wave functions of the binding energy and dissociation temperature of charmonium and bottomonium in the background of the baryonic chemical potential. Omugbe et al. \cite{r33} employed the NU method to find an approximate solution to the fractional SE with a spin-spin-dependent Cornell potential, with improved accuracy in the energy spectra of mass spectra of bottomonium, charmonium, and bottom-charm systems. Abu-Shady and Fath-Allah \cite{r34} have systematically studied the integration of the effects of topological defects by solving the radial SEs of heavy quarkonia in cosmic string spacetime with an extended Cornell potential using the GFD-ENU method. They obtained both classical and fractional  paradigm energy eigenvalues and masses, and explained how the spacetime topology affects the hadron spectroscopy. Abu-Shady et al. \cite{r35} did a thorough study of the medium effects and the dynamics of dissociation. The study examined the effect of anisotropic plasma on dissociation of bottomonium and discovered that topological defects and plasma anisotropy enhance binding energies.The framework has recently been extended to cover more complex systems and environments. Abu-Shady and Fath-Allah \cite{r36} used the GFNU method to study the charmonium and bottomonium binding energies in an anisotropic-dense QGP with topological defects, to elucidate the competing contributions of plasma anisotropy, temperature, chemical potential and fractional derivatives to binding energy and patterns of state splitting. In Ref. \cite{r37}, the authors studied spin-spin interactions and fractional-order effects on fully-heavy tetraquark masses in topological defect space-times, using the GFNU method to calculate binding energies using an extended Cornell potential, which provided predictions of ground and excited states which are close to those predicted by theory.

More recently Ongedo et al. \cite{r38} used a series of fractional derivatives to formulate a SE which includes Cornell and harmonic potentials and Gaussian-approximated spin variables, resulting in the solution of the Heun functions which address spin degeneracies in various quark configurations. By determining the masses of heavy pentaquarks with spin-spin interactions in a fractal space-time with a global monopole, they showed the versatility of the fractional derivative methods to analyze unusual hadron configurations. Combined, the sequential developments provide a mathematically sound and physically informative basis on which to study hadronic systems in complex situations. They are more predictive of mass spectra, binding dynamics, and thermodynamic properties than more conventional integer-order quantum mechanical methods, and effectively capture the effects of topo
logical defects, plasma anisotropy and fractal spacetime structures, which are increasingly being seen as key components of true QCD phenomenology. The sequential developments show that the GFD methods are a mathematically sound and physically meaningful framework in which to model hadronic systems in complex environments and which provide better predictive power of mass spectra, binding dynamics and thermodynamic properties than do traditional integer-order quantum mechanical approaches, and which seamlessly incorporate the effects of topological defects, plasma anisotropy and fractal spacetime structures, which are increasingly being recognized as critical factors in the realistic phenomenology of QCD.

Despite significant advances in the use of the GFD framework to model hadronic systems, such as heavy quarkonia in quark-gluon plasma and tetraquarks in vacuum, a comprehensive analysis of exotic multiquark states in extreme thermal environments remains incomplete. Particularly, the combined effects of plasma anisotropy and fractional-order dynamics on the thermal dissociation and mass spectra of four-quark systems have yet to be thoroughly investigated. To fill this gap, this paper aims to investigate the mass spectra, binding energies, and thermal dissociation properties of fully-heavy ($cc\bar c \bar c$ and $bb\bar b \bar b$) and hidden-heavy ($c\bar q c\bar q$ and $b\bar qb\bar q$) tetraquarks within the framework of an anisotropic hot QCD medium. This study uses the GFD framework in conjunction with the PGFNU scheme to elucidate how fractional-order parameters and plasma anisotropy uniquely influence the stability and spectroscopic properties of these exotic hadrons, and compares the results to predictions from conventional integer-order quantum mechanical models.
\\
The paper will be structured in the following way: Section 2 will introduce the GFNU method, and Section 3 will outline the parametric form of this method. In Section 4, the bound-state solution of non-relativistic quark model will be presented. Section 5 introduces the discussion, while Section 6 concludes it.
\section{The Generalized Fractional  Nikiforov–Uvarov Method}
A new formulation in fractional calculus is the GFD, which has been suggested as an important improvement on the classical definitions including the Caputo and Riemann–Liouville derivatives. This strategy preserves important analytical features including the product and quotient rules for differentiation and important findings including Rolle's theorem and the mean value theorem. Therefore, the GFD framework provides a mathematical foundation that will make fractional calculus more comprehensive and provide a more coherent and efficient treatment of fractional differential equations, as explained in Ref. \cite{r39}.	For a function \(\psi : (0, \infty) \to \mathbb{R}\), the generalized fractional derivative of order \(0 < \alpha \leq 1\) of \(\psi(s)\) at \(s > 0\) is defined as
	\begin{equation}\label{x1}
		D^{\text{GFD}} \psi(s) = \lim_{\epsilon \to 0} \frac{\psi\left(s + \frac{\Gamma(\beta)}{\Gamma(\beta - \alpha + 1)} \epsilon s^{1-\alpha}\right) - \psi(s)}{\epsilon}; \quad \beta > -1, \quad \beta \in \mathbb{R}^+
	\end{equation}
	The characteristics of the GFD are,
	\begin{align}
		\text{I.} & \quad D^\alpha \left[ \psi_{nl}(s) \right] = \delta \, s^{1-\alpha} \, \dot{\psi}_{nl}(s), \\
		\text{II.} & \quad D^\alpha \left[ D^\alpha \psi(s) \right] = \delta^2 \left[(1-\alpha) \, s^{1-2\alpha} \, \dot{\psi}_{nl}(s) + s^{2-2\alpha} \, \psi_{nl}''(s)\right],
	\end{align}
	where, $\delta = \frac{\Gamma[\beta]}{\Gamma[\beta - \alpha + 1]}, \quad \text{with } 0 < \alpha \leq 1, \quad 0 < \beta \leq 1.$
	By using the GFD, the PGFNU method is obtained. The second-order parametric generalized differential equation can be solved in the fractional manner as shown in Ref. \cite{r40}. 
	\begin{equation}\label{x4}
		D^\alpha[ D^\alpha \psi(s) ] + \frac{\bar{\tau}(s)}{\sigma(s)} D^\alpha \psi(s) + \frac{\bar{\sigma}(s)}{\sigma^2(s)} \psi(s) = 0,	
	\end{equation}
	where \(\bar{\sigma}(s), \sigma(s)\) and \(\bar{\tau}(s)\) are polynomials of \(2\alpha\)--th, \(2\alpha\)--th and \(\alpha\)--th degree.
The following transformation is used to obtain the appropriate solution for Eq. (\ref{x4}) by separation of variables. 
	\begin{equation}\label{x5}
		\psi(s) = \Phi(s) y(s),	
	\end{equation}
Eq. (\ref{x4}) can be expressed as in Ref. \cite{r36}.
	\begin{equation}\label{x6}
		\sigma(s) D^\alpha [D^\alpha y(s)] + \tau(s) D^\alpha y(s) + \lambda y(s) = 0,	
	\end{equation}
	where,
	\begin{equation}\label{x7}
		\sigma(s) = \pi(s) \frac{\Phi(s)}{D^\alpha \Phi(s)},	
	\end{equation}
	and
	\begin{equation}\label{x8}
		\pi(s) = \frac{D^\alpha \sigma(s) - \bar{\tau}(s)}{2} \pm \sqrt{\left( \frac{D^\alpha \sigma(s) - \bar{\tau}(s)}{2} - \bar{\sigma}(s) + K \sigma(s) \right)^2},	
	\end{equation}
where $\pi(s)$  is the first degree polynomial.
The values of $K$ in the square-root term of Eq. (\ref{x8}) can be chosen so that the radicand becomes a perfect square. This condition is satisfied when the discriminant of the expression is equal to zero.	 
	\begin{equation}\label{x9}
		\lambda = K + D^\alpha \pi(s),
	\end{equation}
where \(\lambda\) is a constant. Inserting \(K\) into Eq. (\ref{x8}), we define
	\begin{equation}\label{x10}
		\tau(s) = \bar{\tau}(s) + 2\pi(s),
	\end{equation}
Given that both $\rho(s) > 0$ and $\sigma(s) > 0$, the derivative of $\tau$ must be negative \cite{r42}. The modified eigenvalue equation then turns into

	\begin{equation}\label{x11}
		\lambda = \lambda_n = -D^\alpha \tau - \frac{n(n-1)}{2} D^\alpha [D^\alpha \sigma(s)].
	\end{equation}
Eq. (\ref{x4}) has a solution which is the product of two independent parts, where \(y(s) = y_n(s)\) is an \(n\)-th degree polynomial which fulfills the kind of the hypergeometric equation.
	\begin{equation}\label{x12}
		y_n(s) = \frac{G_n}{\rho(s)} (D^\alpha)^n (\sigma(s)^n \rho_n(s)),
	\end{equation}
where \( G_n \) is a constant for normalization, and \( \rho(s) \) is a weight function that satisfies the equation below.
	\begin{equation}\label{x13}
		D^\alpha \omega(s) = \frac{\tau(s)}{\sigma(s)} \omega(s), \quad \omega(s) = \sigma(s) \rho(s),
	\end{equation}
	\begin{equation}\label{x14}
		D^\alpha [\sigma(s) \rho(s)] = \tau(s) \sigma(s).
	\end{equation}
	\section{ The Parametric GFNU Method}
The SE reduces to a second-order parametric generalized differential equation \cite{r43} given by:
	\begin{equation}\label{x15}
		D^\alpha [D^\alpha \psi(s)] + \frac{p_1 - p_2 s^\alpha}{s^\alpha (1 - p_3 s^\alpha)} D^\alpha \psi(s) + \frac{-\xi_1 s^{2\alpha} + \xi_2 s^\alpha - \xi_3}{(s^\alpha (1 - p_3 s^\alpha))^2} \psi(s) = 0.
	\end{equation}
	\begin{equation}\label{x16}
		\bar{\tau}(s) = p_1 - p_2 s^\alpha,
	\end{equation}
	\begin{equation}\label{x17}
		\sigma(s) = s^\alpha (1 - p_3 s^\alpha),
	\end{equation}
	\begin{equation}\label{x18}
		\tilde{\sigma}(s) = -\xi_1 s^{2\alpha} + \xi_2 s^\alpha - \xi_3.
	\end{equation}
	Altering them in Eq. (\ref{x8}), we get:
	\begin{equation}\label{x19}
		\pi = p_4 + p_5 s^\alpha \pm \sqrt{(p_6 - K p_3) s^{2\alpha} + (p_7 + K) s^\alpha + p_8},
	\end{equation}
	where,
	\begin{equation}\label{x20}
		p_4 = \frac{1}{2} (\alpha - p_1),
	\end{equation}
	\begin{equation}\label{x21}
		p_5 = \frac{1}{2} (p_2 - 2p_3 \alpha),
	\end{equation}
	\begin{equation}\label{x22}
		p_6 = p_5^2 + \xi_1,
	\end{equation}
	\begin{equation}\label{x23}
		p_7 = 2p_4p_5 - \xi_2,
	\end{equation}
	\begin{equation}\label{x24}
		p_8 = p_4^{2} + \xi_3.
	\end{equation}
The NU method requires the radicand in Eq. (\ref{x19}) to be a perfect square of a polynomial., so that
	\begin{equation}\label{x25}
		K = -(p_7 + 2p_3p_8) \pm 2\sqrt{p_8p_9},
	\end{equation}
	where,
	\begin{equation}\label{x26}
		p_9 = p_3 p_7 + p_3^2 p_8 + p_6.
	\end{equation}
The negative values of \( K \) are obtained by:
	\begin{equation}\label{x27}
		K = -(p_7 + 2 p_3 p_8) - 2 \sqrt{p_8 p_9}
	\end{equation}
	So that \(\pi\) becomes
	\begin{equation}\label{x28}
		\pi = p_4 + p_5 s^\alpha - [(\sqrt{p_9} + p_3 \sqrt{p_8}) s^\alpha - \sqrt{p_8}]
	\end{equation}
	From Eqs. (\ref{x10}), (\ref{x16}) and (\ref{x28}), we get
	\begin{equation}\label{x29}
		\tau = p_1 + 2 p_4 - (p_2 - 2 p_5) s^\alpha - [(\sqrt{p_9} + p_3 \sqrt{p_8}) s^\alpha - \sqrt{p_8}]
	\end{equation}
	From Eqs. (\ref{x10}) and (\ref{x26}), we get,
	\begin{equation}\label{x30}
		\begin{split}
			D^\alpha \tau = \delta [- \alpha (p_2 - 2 p_5) - 2 \alpha (\sqrt{p_9} + p_3 \sqrt{p_8})]
			\\
			= \delta [- 2 \alpha^2 p_3 - 2 \alpha (\sqrt{p_9} + p_3 \sqrt{p_8})] < 0
		\end{split}
	\end{equation}
	Eqs. (\ref{x9}) and (\ref{x11}) provide the energy spectrum equation as follows:
	\begin{equation}\label{x31}
		n \delta \alpha p_2 - (2n+1) \delta \alpha p_5 + (2n+1) \delta \alpha (\sqrt{p_9} + p_3 \sqrt{p_8}) 
		+ n(n-1) \delta^2 \alpha^2 p_3 + p_7 + 2 p_3 p_8 + 2 \sqrt{p_8 p_9} = 0
	\end{equation}
Setting $\alpha = \beta  = 1$ reduces the expression to the classical energy eigenvalue reported in Ref. \cite{r44}.	
	\begin{equation}\label{x32}
		n p_2 - (2n+1) p_5 + (2n+1) (\sqrt{p_9} + p_3 \sqrt{p_8})
		+ n(n-1) p_3 + p_7 + 2 p_3 p_8 + 2 \sqrt{p_8 p_9} = 0
	\end{equation}
From Eq. (\ref{x14}), we obtain
	\begin{equation}\label{x33}
		\rho(s) = s^{\frac{p_{10} - \alpha}{\delta}} \left( 1 - p_3 s^{\alpha} \right)^{\frac{p_{11}}{\alpha \delta p_3} \frac{p_{10}}{\alpha \delta} \frac{1}{\delta}}
	\end{equation}
Eq. (\ref{x12}) yields:
	\begin{equation}\label{x34}
		y_n(s) = P_n^{ \left( \frac{p_{10} - \alpha}{\delta}, \frac{p_{11}}{\alpha \delta p_3} \frac{p_{10}}{\alpha \delta} \frac{1}{\delta} \right) } \left( 1 - 2 p_3 s^{\alpha} \right)
	\end{equation}
	where,
	\begin{equation}\label{x35}
		p_{10} = p_1 + 2 p_4 + 2 \sqrt{p_8}
	\end{equation}
	\begin{equation}\label{x36}
		p_{11} = p_2 - 2p_5 + 2\left(\sqrt{p_9} + p_3\sqrt{p_8}\right)
	\end{equation}
Eq. (\ref{x7}) yields the general wave function solution.
	\begin{equation}\label{x37}
		\psi(s) = s^{\frac{p_{12}}{\alpha \delta}} (1 - p_3 s^{\alpha})^{\frac{-p_{13}}{\alpha \delta p_3}} P_n^{\left(\frac{p_{10}-\alpha}{\alpha \delta}, \frac{p_{11}}{\alpha \delta p_3} - \frac{p_{10}}{\alpha \delta} - \frac{1}{\delta}\right)} (1 - 2p_3 s^{\alpha})
	\end{equation}
	where \(P_n^{(v_1, v_2)}\) are Jacobi polynomials.
	\begin{equation}\label{x38}
		p_{12} = p_4 + \sqrt{p_8}
	\end{equation}
	\begin{equation}\label{x39}
		p_{13} = p_5 - \left(\sqrt{p_9} + p_3\sqrt{p_8}\right)
	\end{equation}
	Some problems, in case \(p_3 = 0\):
	\begin{equation}\label{x40}
		\lim_{p_3 \to 0} P_n^{\left(\frac{p_{10}-\alpha}{\alpha \delta}, \frac{p_{11}}{\alpha \delta p_3} - \frac{p_{10}}{\alpha \delta} - \frac{1}{\delta}\right)} (1 - p_3 s^{\alpha}) = L_n^{\left(\frac{p_{10}-\alpha}{\alpha \delta}, \frac{p_{11}}{\alpha \delta p_3} - \frac{p_{10}}{\alpha \delta} - \frac{1}{\delta}\right)}(e^{\alpha \delta} s^{\alpha})
	\end{equation}
	\begin{equation}\label{x41}
		\lim_{p_3 \to 0} (1 - p_3 s^{\alpha})^{\frac{-p_{13}}{\alpha \delta p_3}} = e^{p_{13} s^{\alpha}}
	\end{equation}
	Eq. (\ref{x37}) becomes
	\begin{equation}\label{x42}
		\psi(s) = s^{\frac{p_{12}}{\alpha \delta}} e^{p_{13} s^{\alpha}} L_n^{\left(\frac{p_{10}-\alpha}{\alpha \delta}, \frac{p_{11}}{\alpha \delta p_3} - \frac{p_{10}}{\alpha \delta} - \frac{1}{\delta}\right)}(e^{\alpha \delta} s^{\alpha})
	\end{equation}
	where \(L_n\) are the Laguerre polynomials, and the solution of Eq. (\ref{x27}) becomes:
	\begin{equation}\label{x43}
		K = -\left(p_7 + 2p_3p_8\right) + 2\sqrt{p_8p_9}
	\end{equation}
The wave function reads
	\begin{equation}\label{x44}
		\psi(s) = s^{\frac{p_{12}^*}{\alpha \delta}} \left( 1 - p_3 s^{\alpha} \right)^{\frac{-p_{13}^*}{\alpha \delta p_3}} P_n^{\left( \frac{p_{10}^* - \alpha}{\alpha \delta}, \frac{p_{11}^*}{\alpha \delta p_3} - \frac{p_{10}^*}{\alpha \delta} - \frac{1}{\delta} \right)} (1 - 2p_3 s^{\alpha})
	\end{equation}
The general form of the energy eigenvalue becomes
	\begin{equation}\label{x45}
		\begin{aligned}
			n \delta & \alpha p_2 - 2n \delta \alpha p_5 + (2n+1)\delta \alpha (\sqrt{p_9} - p_3 \sqrt{p_8}) + n(n-1)\delta^2 \alpha^2 p_3 + p_7 \\
			& + 2p_3 p_8 - 2\sqrt{p_8p_9} + \delta \alpha p_5 = 0
		\end{aligned}
	\end{equation}
	where,
	\begin{equation}\label{x46}
		p_{10}^* = p_1 + 2p_4 - 2\sqrt{p_8}
	\end{equation}
	\begin{equation}\label{x47}
		p_{11}^* = p_2 - 2p_5 + 2(\sqrt{p_9} - p_3 \sqrt{p_8})
	\end{equation}
	\begin{equation}\label{x48}
		p_{12}^* = p_4 - \sqrt{p_8}
	\end{equation}
	\begin{equation}\label{x49}
		p_{13}^* = p_5 - (\sqrt{p_9} - p_3 \sqrt{p_8})
	\end{equation}
	\section{The non-relativistic quark model}
	We begin by establishing the framework for the quark model employed in this study. Considering that the systems under investigation consist of quarkonia $(Q\bar Q)$ and all-heavy tetraquarks ($T_{4Q}$) in a diquark–antidiquark configuration $([QQ][\bar Q\bar Q])$, we operate under the assumption that both systems can be effectively reduced to a two-body problem. It makes sense to employ a non-relativistic framework with static potentials since the constituents' momentum should be extremely small in relation to their rest mass. In this regard, we use the time-independent SE with spherically symmetric potentials, where the radial part is expressed as follows:
	\begin{equation}\label{x50}
		\left[ \frac{d^2}{dr^2} + \frac{N-1}{r} \frac{d}{dr} - \frac{l(l+N-2)}{r^2} + 2M (E - V(r)) \right] \Psi(r) = 0,
	\end{equation}
where $l$ is the angular momentum quantum number, $N$ is the dimensionality, and $M$ is the reduced mass. Putting the wave function 
	$\Psi(r) = r^{\frac{1-N}{2}} R(r)
	$.
This yields the following SE:
	\begin{equation}\label{x51}
		\left[ \frac{d^2}{dr^2} + 2M (E - V(r)) - \frac{\left(l+\frac{N-2}{2}\right)^2 - \frac{1}{4}}{r^2} \right] R(r) = 0.
	\end{equation}
According to Ref. \cite{r24} and its references, the potential function is defined as:
	\begin{equation}\label{x52}
		V(r) = -\frac{a}{r} (1 + \mu r) e^{-\mu r} + \frac{2\sigma}{\mu} (1 - e^{-\mu r}) - \sigma r e^{-\mu r},
	\end{equation}
	where,
	\begin{equation}\label{x53}
		\left( \frac{\mu}{m_D} \right)^{-4} = 1 + \xi \left(a - \frac{2^b (a-1)+(1+\xi)^{\frac{1}{8}}}{(3+\xi)^b}\right) \left(1 + \frac{c(\theta)(1+\xi)^d}{(1+e\xi^2)}\right),
	\end{equation}
	where $\xi$ denotes to the anisotropic parameter, and the function \( c(\theta) \) is obtained by imposing the small-\(\xi\) limit.
	\begin{equation}\label{x54}
		c(\theta) = \frac{3\pi^2 \cos(2\theta)(9+4\sqrt{3}-4\sqrt{6})\pi^2 + 64(\sqrt{6}-3)}{4(\sqrt{3}(\sqrt{2}-1)\pi^2 - 16(\sqrt{6}-3))}
	\end{equation}
We adopt the parameter values \( a = 0.385 \), \( b = \frac{1}{2} \), \( d = \frac{3}{2} \), and \( e = \frac{1}{3} \), following Ref.~\cite{r45}. The analysis is carried out for \( \theta = 0^\circ \) and \( \theta = 90^\circ \), where \( \theta = 0^\circ \) corresponds to the particle momentum aligned with the anisotropy direction, while \( \theta = 90^\circ \) represents momentum perpendicular to it. In the isotropic limit (\( \xi = 0 \)), the screening mass reduces to \( \mu(\theta, \xi, T) = m_D(T) \), where \( m_D(T) \) denotes the temperature-dependent Debye mass.
	\begin{equation}\label{x55}
		m_D = A \, g \, T \, \sqrt{\frac{N_c}{3} + \frac{N_f}{6}},
	\end{equation}
where \( N_c \) and \( N_f \) denote the number of colors and flavors, respectively. We use \( \sigma = 0.223 \, \text{GeV}^2 \), following Ref.~\cite{r45}. In Eq.~(\ref{x52}), the exponential term \( e^{-\mu r} \) is expanded in the limit \( \mu r \ll 1 \), as discussed in Ref.~\cite{r46}. Consequently, Eq.~(\ref{x52}) can be expressed as
	\begin{equation}\label{x56}
		V(r) = A_1 r^3 + A_2 r^2 + A_3 r + \frac{A_4}{r},
	\end{equation}
	where,
	\begin{equation}\label{x57}
		A_1 = -\frac{1}{2} \sigma \mu^2,
	\end{equation}
	\begin{equation}\label{x58}
		A_2 = -\frac{1}{2} a \mu^3,
	\end{equation}
	\begin{equation}\label{x59}
		A_3 = \frac{1}{2} a \mu^2 + \sigma,
	\end{equation}
	\begin{equation}\label{x60}
		A_4 = -a.
	\end{equation}
	The radial SE with the interaction described by the extended Cornell potential as defined in Ref.~\cite{r47}, and using the transformation \( z = e^{-\gamma r} \), takes the form
	\begin{equation}\label{x61}
		\frac{d^2R}{dr^2} + \frac{1}{z} \frac{dR}{dr} - \frac{1}{z^2(1-z)^2} \{-\xi_1 z^2 + \xi_2 z - \xi_3\} R(z) = 0,
	\end{equation}
	where
	\begin{equation}\label{x62}
		\xi_1 = -\frac{2ME}{\gamma^2} + B_1, \quad \xi_2 = -\frac{4ME}{\gamma^2} + B_2, \quad \xi_3 = -\frac{2ME}{\gamma^2} + B_3,
	\end{equation}
	with
	\begin{equation}\label{x63}
		B_1 = \frac{20MA_1}{\gamma^5} + \frac{12MA_2}{\gamma^4} + \frac{6MA_3}{\gamma^3},
	\end{equation}
	\begin{equation}\label{x64}
		B_2 = \frac{10MA_1}{\gamma^5} + \frac{6MA_2}{\gamma^4} + \frac{2MA_3}{\gamma^3} + \frac{2Ma_4}{\gamma},
	\end{equation}
	\begin{equation}\label{x65}
		B_3 = \frac{2MA_1}{\gamma^5} + \frac{2MA_2}{\gamma^4} + \frac{2MA_3}{\gamma^3} + \frac{2Ma_4}{\gamma} + \left(l+\frac{N-2}{2}\right)^2 - \frac{1}{4},
	\end{equation}
The generalized fractional part of the SE as
	\begin{equation}\label{x66}
		D^\alpha \left[ D^\alpha R(z) \right] + \frac{1-z^\alpha}{z^\alpha (1-z^\alpha)} D^\alpha R(z) + \frac{-\xi_1 z^{2\alpha} + \xi_2 z^{\alpha} - \xi_3}{(z^\alpha (1-z^\alpha))^2} R(z) = 0,
	\end{equation}
By using the following coefficients:
	\begin{equation}\label{x68}
		p_1 = 1, \quad p_2 = 1, \quad p_3 = 1, \quad p_4 = \frac{1}{2} (\delta \alpha - 1),
	\end{equation}
	\begin{equation}\label{x69}
		p_5 = \frac{1}{2} (1 - 2\delta \alpha), \quad p_6 = \frac{1}{4} (1 - 2\delta \alpha)^2 - \frac{2M E}{\gamma^2} + B_1,
	\end{equation}
	\begin{equation}\label{x70}
		p_7 = \frac{1}{2} (\delta \alpha - 1)(1 - 2\delta \alpha) + \frac{4M E}{\gamma^2} - B_2,
	\end{equation}
	\begin{equation}\label{x71}
		p_8 = \frac{1}{4} (\delta \alpha - 1)^2 - \frac{2M E}{\gamma^2} + B_3, \quad p_9 = \frac{1}{4} \delta^2 \alpha^2 + B_1 - B_2 + B_3,
	\end{equation}
	\begin{equation}\label{x72}
		p_{10} = \delta \alpha + 2 \sqrt{\frac{1}{4} (\delta \alpha - 1)^2 - \frac{2M E}{\gamma^2} + B_3},
	\end{equation}
	\begin{equation}\label{x73}
		p_{11} = 2 \delta \alpha + 2 \left( \sqrt{\frac{1}{4} \delta^2 \alpha^2 + B_1 - B_2 + B_3} + \sqrt{\frac{1}{4} (\delta \alpha - 1)^2 - \frac{2M E}{\gamma^2} + B_3} \right),
	\end{equation}
	\begin{equation}\label{x74}
		p_{12} = \frac{1}{2} (\delta \alpha - 1) + \sqrt{\frac{1}{4} (\delta \alpha - 1)^2 - \frac{2M E}{\gamma^2} + B_3},
	\end{equation}
	\begin{equation}\label{x75}
		p_{13} = \frac{1}{2}(1 - 2\delta\alpha) - \left( \sqrt{\frac{1}{4}\delta^2\alpha^2 + B_1 - B_2 + B_3} + \sqrt{\frac{1}{4}(\delta\alpha - 1)^2 - \frac{2ME}{\gamma^2} + B_3} \right),
	\end{equation}
The generalized fractional energies are given by:
	\begin{equation}\label{x76}
		E = \frac{\gamma^2}{2M}\left(B_3 + \frac{1}{4}(\delta\alpha - 1)^2\right) - \frac{\gamma^2}{2M}\left(\frac{B_1 - B_3 - \left((n+\frac{1}{2})\delta\alpha + \sqrt{\frac{1}{4}\delta^2\alpha^2 + B_1 - B_2 + B_3}\right)^2}{2\left((n+\frac{1}{2})\delta\alpha + \sqrt{\frac{1}{4}\delta^2\alpha^2 + B_1 - B_2 + B_3}\right)}\right).
	\end{equation}
The generalized fractional wave function is
	\begin{equation}\label{x77}
		\begin{aligned}
			R(z) &= N z^{\frac{1}{2}(\delta\alpha - 1) + \sqrt{\frac{1}{4}(\delta\alpha - 1)^2 - \frac{2ME}{\gamma^2} + B_3}} (1 + z^\alpha)^{\frac{1}{2}\delta\alpha + \sqrt{\frac{1}{4}\delta^2\alpha^2 + B_1 - B_2 + B_3}} \\
			&\quad \times (1 - z^\alpha)^{-\alpha + \delta\alpha + 2\sqrt{\frac{1}{4}(\delta\alpha - 1)^2 - \frac{2ME}{\gamma^2} + B_3}} \cdot {_2F_1}(a,b;c;z^\alpha),
		\end{aligned}
	\end{equation}
where \( N \) is the constant of normalization.
\section{Results and Discussion}
In this part, we analyze the tetraquarks properties in the medium by employing the PGFNU method to solve the radial SE and find the dissociation energies for 1\textit{p} states. we define the dissociation energy $E_b(T)$ as Refs. \cite{r48}-\cite{r51}:
	\begin{equation}\label{x78}
		E_b(T)=V(r\to \infty)-E(T),
	\end{equation}
The dissolution of the bound state occurs at the dissociation temperature ($T_{\text{dis}}$), which corresponds to the temperature where the binding energy drops to zero. The binding energy is extracted from the temperature-dependent energy eigenvalue \( E(T) \), found by solving the SE with the Cornell potential replaced by a temperature-dependent potential \( V(r,T) \). Thermal effects are incorporated through the Debye screening mass appearing in the in-medium potential.		
Two distinct cases are considered in this work: the conventional paradigm, corresponding to \( \alpha = \beta = 1 \), and the fractional  paradigm, for which \( \alpha, \beta < 1 \).
Within the framework of fractional quantum mechanics, the fractional parameter (\( 0 < \alpha \leq 1 \)) characterizes the intrinsic properties of the quantum system. The obtained results indicate that the internal dynamics of the system become more pronounced for smaller values of \( \alpha \), while these effects gradually diminish as \( \alpha \) approaches the classical limit (\( \alpha = 1 \)). Moreover, the energy of the system decreases for smaller fractional parameters, implying that the bound states become more stable and more strongly bound in the fractional  paradigm compared to the standard quantum-mechanical case.
The present study primarily focuses on the influence of \( \xi \) on the binding energy. It is observed that raising \( \xi \) strengthens the interaction potential caused by the enhancement of the effective screening mass, which consequently increases the dissociation energy compared to the isotropic medium. In contrast, increasing the temperature lowers the interaction potential and reduces the binding energy.

Figs. \textbf{1(a)} and \textbf{1(b)} illustrate the dissociation energy of the  $cc\bar c \bar c$ \( p \)-state as a function of temperature for both the fractional  paradigm (\( \alpha = \beta = 0.01 \)) and the conventional paradigm (\( \alpha = \beta = 1 \)) at \( \theta = 0 \), for different values of \( \xi \). The results indicate that an increase in temperature causes a decrease in the dissociation energy in all cases, consistent with previous studies. In contrast, increasing \( \xi \) implies to an enhancement of the dissociation energy, indicating stronger binding in the anisotropic medium.
Similarly, Figs. \textbf{1(c)} and \textbf{1(d)} exhibit the same behavior at \( \theta = 90^\circ \). The dissociation energy continues to decrease with increasing temperature; however, its magnitude is smaller compared to the case of \( \theta = 0^\circ \).

Figs. \textbf{2(a)} and \textbf{2(b)} display the change of the dissociation energy with temperature for the $bb\bar b \bar b$ p-state. The same general behavior is observed, where the dissociation energy decreases with increasing temperature and increases with larger values of $\xi$. Comparing the  $cc\bar c \bar c$ and $bb\bar b \bar b$ systems, it is found that the dissociation energy of the  $cc\bar c \bar c$ state is greater than that of the $bb\bar b \bar b$ state, 
which aligns with previous findings  published in the literature.
\subsection{Dissociation Temperature}
The dissociation temperature ($T_D$)  can be derived using the condition that the binding energy becomes comparable to the thermal energy of the medium $E_b(T) \approx T_D$,
which means that the bound state melts when thermal fluctuations become sufficiently strong to overcome the binding energy. However, this criterion takes into account only the real part of the binding energy and neglects medium-induced decay effects. Therefore, the $T_D$ obtained from the binding-energy criterion may differ from that determined using the thermal-width condition $\Gamma (T_D) \geq$ 2 Re $  E_b(T_D)$, where the dissociation occurs when the thermal width becomes larger than twice the real part of the binding energy.

In the current investigation, the effects of isotropic and
anisotropic media on the $T_D$ of both
$cc\bar c \bar c$ and $bb\bar b \bar b$ states are
investigated by taking $T_c=0.17$ as the critical
temperature \cite{r52}.
 The results presented in Table (\ref{tab:t1}) show that the $(T_D)$ increases in the anisotropic medium compared with the isotropic case. This behavior is directly related to the enhancement of the binding energy, which makes the bound states more stable in the anisotropic medium. 
For the fractional  paradigm with $\alpha=0.01, \theta=0$, the $cc\bar c \bar c$ state dissociates at $0.3564T_c$ in the isotropic medium, while in the anisotropic medium with  $\xi= 0.1$, the $T_D$ become $0.3564T_c$, respectively. At $\alpha=0.02$, the corresponding $T_D$ are 0.3570$T_c$, and 0.3707$T_c$. For $\alpha=0.03$, the $T_D$ become 0.3577$T_c$, 0.3620$T_c$, and 0.3714$T_c$, respectively.
In the conventional paradigm $\alpha=1$, $\theta=0$, the $cc\bar c \bar c$ state dissociates at 0.4149$T_c$ in the isotropic medium, while the $T_D$ increase to 0.4195$T_c$ and 0.4304$T_c$ for $\xi= 0.1$ and $\xi = 0.3$, respectively. For the case $\theta=90$, similar behavior is observed.
In the fractional  paradigm with $\alpha=0.01$, the $cc\bar c \bar c$ state dissociates at 0.3564$T_c$ in the isotropic medium, while the $T_D$ become 0.3589$T_c$ and 0.3637$T_c$ for $\xi= 0.1$ and $\xi = 0.3$, respectively. Similar trends are obtained for $\alpha=0.02$ and $\alpha=0.03$. In the conventional paradigm ($\alpha=1$), the $T_D$ are 0.4145$T_c$, 0.4174$T_c$, and 0.4230$T_c$ for $\xi= 0.1$ and $\xi = 0.3$, respectively.

Table (\ref{tab:t2}) presents the corresponding results for the $bb\bar b \bar b$ tetraquark state. In the fractional  paradigm with $\alpha=0.01$ and $\theta=0$, the $bb\bar b \bar b$ state dissociates at 0.4038$T_c$ in the isotropic medium, while the $T_D$ increase to 0.4088$T_c$ and 0.4193$T_c$ for $\xi= 0.1$ and $\xi= 0.3$, respectively. Similar results are obtained for $\alpha=0.02$ and $\alpha=0.03$. In the conventional paradigm ($\alpha=1$), the $T_D$ become 0.4217$T_c$, 0.4269$T_c$, and 0.4379$T_c$, respectively.
For $\theta=90$, the same qualitative behavior is maintained. The $T_D$ slightly decrease compared with the case $\theta=0$, but the differences remain relatively small.
The obtained results clearly indicate that the $T_D$ increases with increasing binding energy, reflecting the stronger stability of the tetraquark states inside the anisotropic medium. Moreover, a clear difference is observed between the fractional and conventional paradigms. In contrast, the angular dependence between $\theta=0$ and $\theta=90$ remains weak.	
Unlike the $cc\bar c \bar c$ state, which dissociates slightly above the critical temperature, the $bb\bar b \bar b$ tetraquark can survive inside the quark–gluon plasma at considerably higher temperatures causes by the very large mass of the bottom quark. This observation is consistent with previous studies.
Overall, the fractional parameter $\alpha$  plays an crucial role in determining the properties of the quantum system within the framework of fractional quantum mechanics. The fractional approach modifies the confinement dynamics and leads to noticeable changes in the binding energy and $T_D$ compared with the classical 
\subsection{Masses of Tetraquark}
To investigate the tetraquark masses, the temperature is first set to zero, reducing the interaction potential to the standard Cornell potential. The tetraquark mass is then defined by
	\begin{equation}\label{x79}
		M = m_{QQ} + m_{\bar{Q}\bar{Q}} + E_{[QQ][\bar{Q}\bar{Q}]}	
	\end{equation} 
Table (\ref{tab:t3}) show the calculated masses of the $cc\bar c \bar c$ tetraquark for the 1\textit{s}, 2\textit{s}, 3\textit{s}, 4\textit{s}, 1\textit{p} and  2\textit{p} states within both the conventional paradigm $(\alpha = \beta = 1)$ and the fractional  paradigm $(\alpha = \beta = 0.8)$. The obtained results show good agreement with earlier theoretical studies Refs.  \cite{r9}, \cite{r30}, \cite{r53}. In Ref. \cite{r9}, the tetraquark system was studied using the Cornell potential together with the spin–spin interaction, where the SE was solved numerically. Ref. \cite{r30} employed the generalized fractional extended NU method to find analytical solutions of the SE using a fractional potential that combines the Cornell and harmonic oscillator potentials. In Ref. \cite{r53}, the tetraquark system was investigated within the nonrelativistic Bethe–Salpeter framework by considering logarithmic, harmonic, and linear interaction potentials in addition to the spin–spin interaction.

Table  (\ref{tab:t4}) presents the calculated masses of the $bb\bar b \bar b$ tetraquark for the 1\textit{s}, 2\textit{s}, 3\textit{s}, 4\textit{s}, 1\textit{p} and  2\textit{p} states in both the conventional paradigm $(\alpha = \beta = 1)$ and the fractional  paradigm $(\alpha = \beta = 0.1)$. The obtained values are found to be consistent with the results reported in Refs. \cite{r30}, \cite{r53}.
In Table (\ref{tab:t5}), the masses of the $c\bar q c \bar q$ tetraquark are evaluated for the same excited states using the conventional paradigm $(\alpha = \beta = 1)$ and the fractional  paradigm $(\alpha = \beta = 0.1)$. The calculated masses are in satisfactory agreement with previous investigations Refs. \cite{r16}, \cite{r19}, \cite{r30}. The authors in Ref. \cite{r16} studied heavy tetraquarks containing hidden charm and bottom quarks through the homogeneous Lippmann–Schwinger integral equation in momentum space within the diquark–antidiquark picture. 
In that work, the tetraquark was considered as a two-body system, and a regularized interaction potential was introduced to remove the singular behavior of the confining potential at large distances. On the other hand, the authors in Ref. \cite{r19} investigated heavy tetraquark spectroscopy using a relativistic diquark–antidiquark model via the two-body Klein–Gordon equation with a vector Coulomb-like potential and a scalar linear confining potential. This approach enabled the calculation of both ground and excited tetraquark states.
Finally, Table  (\ref{tab:t6}) lists the masses of the $b\bar q b \bar q$  tetraquark for the 1\textit{s}, 2\textit{s}, 3\textit{s}, 4\textit{s}, 1\textit{p} and  2\textit{p} states using both the conventional paradigm $(\alpha = \beta = 1)$ and the fractional  paradigm $(\alpha = \beta = 0.4)$. The obtained findings are also in acceptable alignment with the previous theoretical investigations given in Refs. \cite{r16}, \cite{r19}, \cite{r30}.

	\begin{figure}[!h]
		\begin{center}
			\begin{minipage}{0.48\textwidth}
				\centering
				\includegraphics[width=1.05\linewidth]{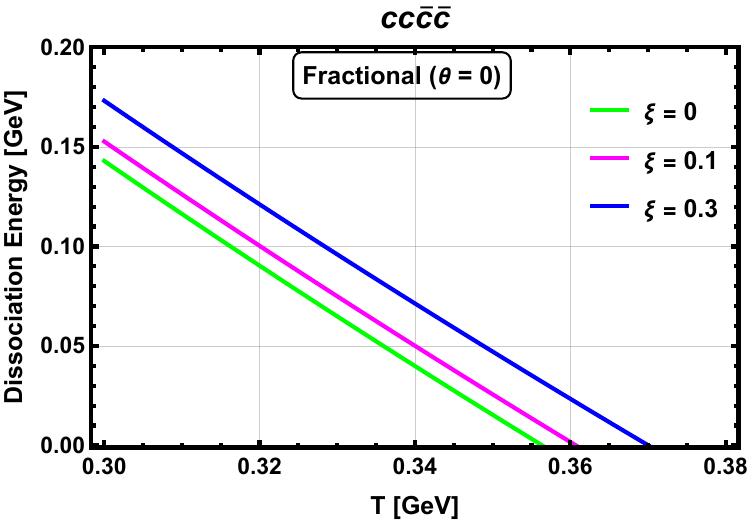}
				\textbf{(a)}
			\end{minipage}
			\quad
			\begin{minipage}{.48\textwidth}
				\centering
				\includegraphics[width=1.05\linewidth]{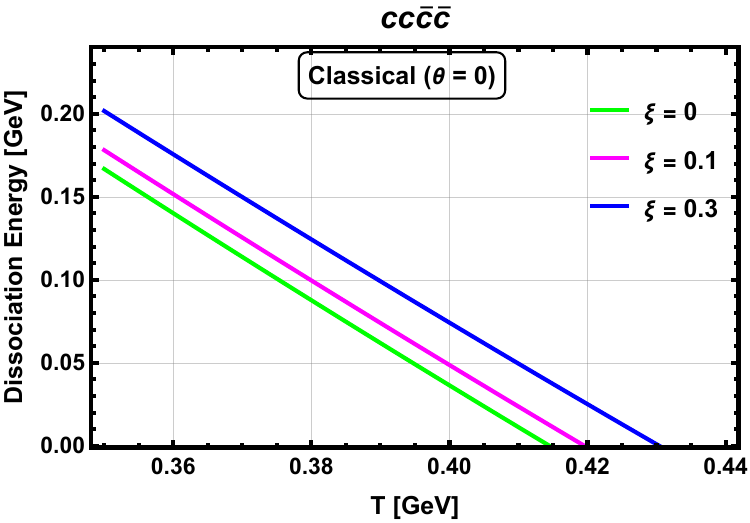}
				\textbf{(b)}
			\end{minipage}
			\quad
			\begin{minipage}{.48\textwidth}
				\centering
				\includegraphics[width=1.05\linewidth]{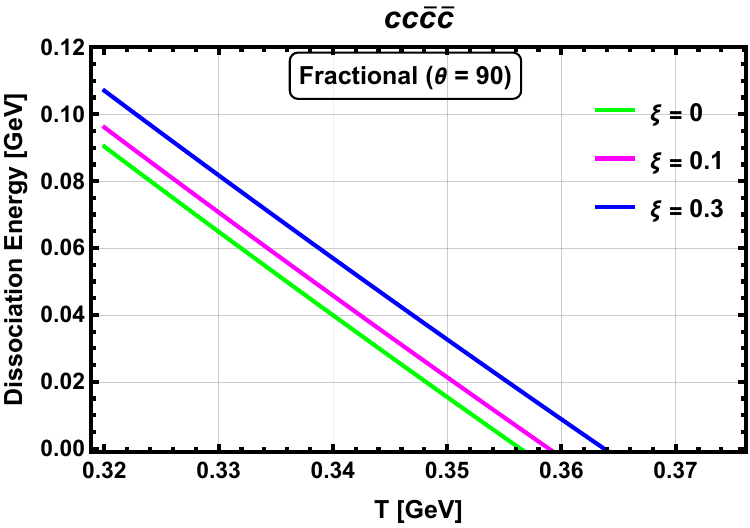}
				\textbf{(c)}
			\end{minipage}
			\quad
			\begin{minipage}{.48\textwidth}
				\centering
				\includegraphics[width=1.05\linewidth]{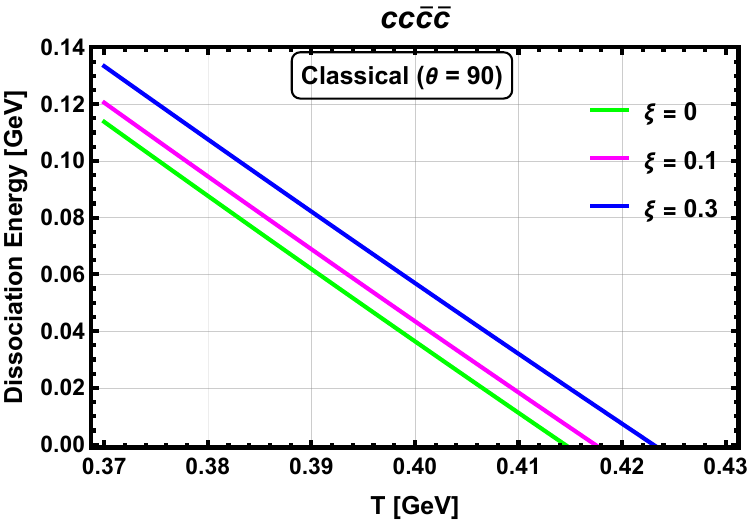}
				\textbf{(d)}
			\end{minipage}
		\end{center}
		\caption{Plot of the dissociation energy of $p$-state of $cc\bar c \bar c$ with $T$ and different values of $\xi$ at (a) $\theta=0$, $\alpha$ = 0.01, (b) $\theta=0$, $\alpha$ = 1, (c) $\theta=90$, $\alpha$ = 0.01 and (d) $\theta=90$, $\alpha$ = 1}\label{f1}
	\end{figure}
	\clearpage
	\begin{figure}[!h]
		\begin{center}
			\begin{minipage}{0.48\textwidth}
				\centering
				\includegraphics[width=1.05\linewidth]{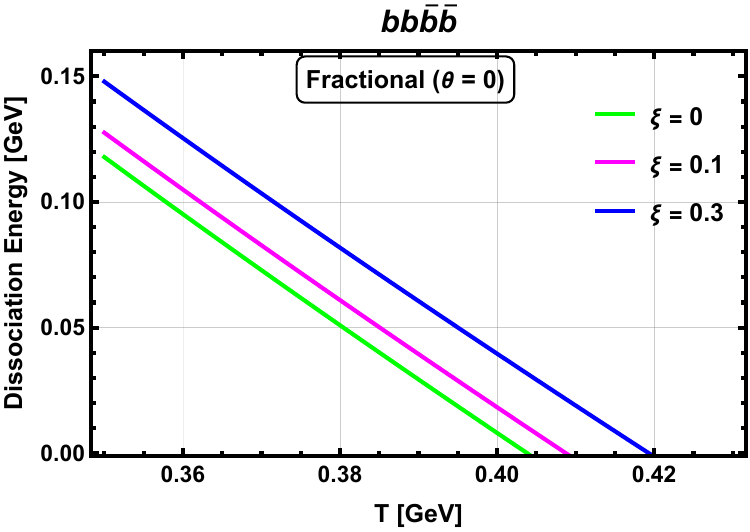}
				\textbf{(a)}
			\end{minipage}
			\quad	
			\begin{minipage}{.48\textwidth}
				\centering
				\includegraphics[width=1.05\linewidth]{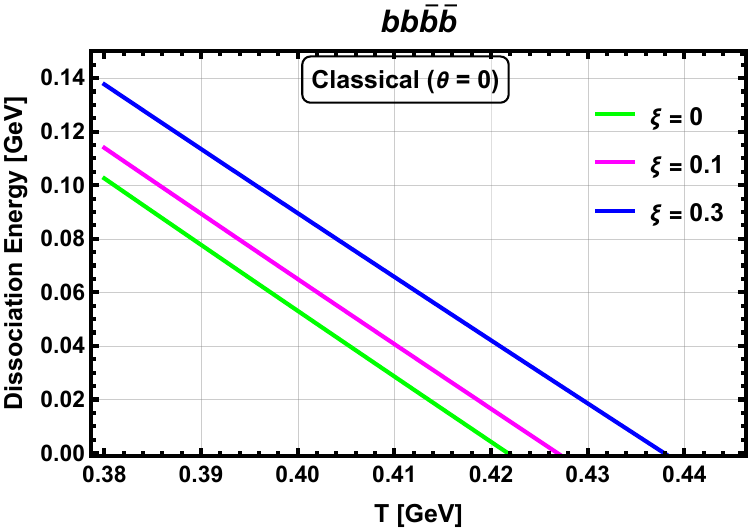}
				\textbf{(b)}
			\end{minipage}
			\quad	
			\begin{minipage}{.48\textwidth}
				\centering
				\includegraphics[width=1.05\linewidth]{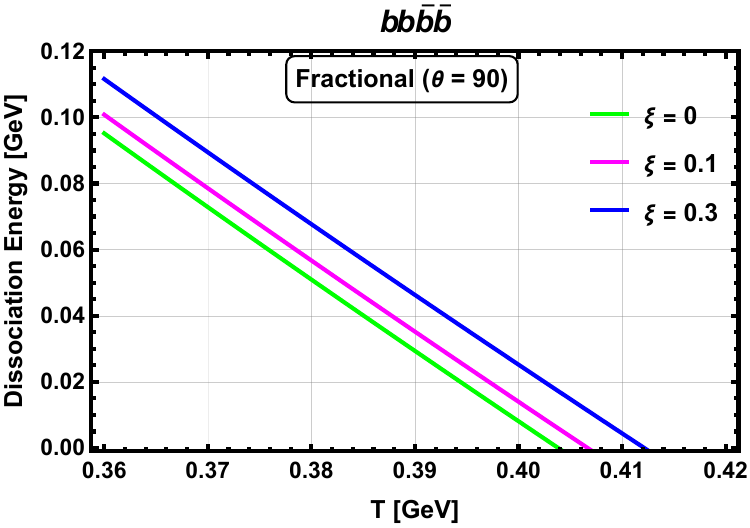}
				\textbf{(c)}
			\end{minipage}
			\quad	
			\begin{minipage}{.48\textwidth}
				\centering
				\includegraphics[width=1.05\linewidth]{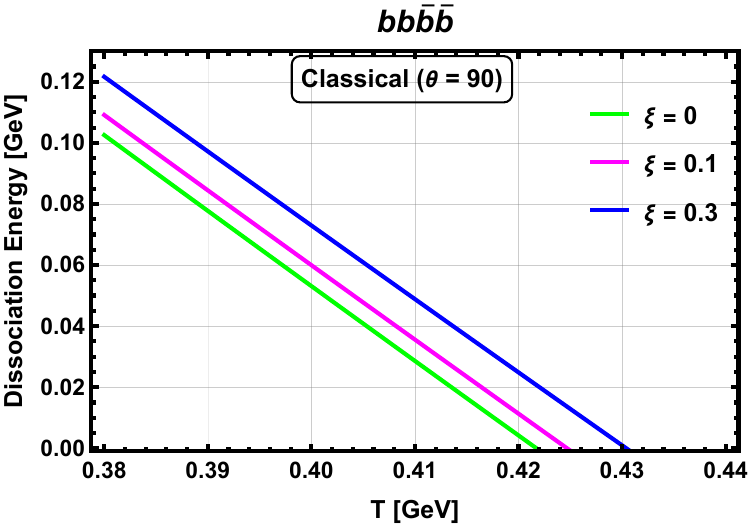}
				\textbf{(d)}
			\end{minipage}
		\end{center}
		\caption{Plot of the dissociation energy of $p$-state of $bb\bar b \bar b$ with $T$ and different values of $\xi$ at (a) $\theta=0$, $\alpha$ = 0.01, (b) $\theta=0$, $\alpha$ = 1, (c) $\theta=90$, $\alpha$ = 0.01 and (d) $\theta=90$, $\alpha$ = 1}\label{f2}
	\end{figure}
	
	\begin{table}[]
		\caption{Dissociation temperature ($T_D$) for
			$cc\bar c \bar c$.}\label{tab:t1}
		\centering
		\renewcommand{\arraystretch}{1.4}
		\begin{tabular}{ccccccc}
			\toprule
			& \multicolumn{3}{c}{$\theta=0$} & \multicolumn{3}{c}{$\theta=90$} \\ 
			\cmidrule(lr){2-4}\cmidrule(lr){5-7}
			$\alpha$ & $\xi=0$      & $\xi=0.1$      & $\xi=0.3$         & $\xi=0$       & $\xi=0.1$       & $\xi=0.3$      \\ 
			\midrule
			0.01                      &   0.3564 $T_c$ & 0.3607 $T_c$ & 0.3701 $T_c$ & 0.3564 $T_c$ & 0.3589 $T_c$ & 0.3637 $T_c$  \\
			0.02                      & 0.3570 $T_c$ & 0.3614 $T_c$ & 0.3707 $T_c$ & 0.3570 $T_c$ & 0.3595 $T_c$ & 0.3644 $T_c$ \\
			0.03                      & 0.3577 $T_c$ & 0.3620 $T_c$ & 0.3714 $T_c$ & 0.3577 $T_c$ & 0.3602 $T_c$ & 0.3650 $T_c$ \\
			1.00                      & 0.4149 $T_c$ & 0.4195 $T_c$ & 0.4304 $T_c$ & 0.4145 $T_c$ & 0.4174 $T_c$ & 0.4230 $T_c$ \\
			\bottomrule
		\end{tabular}
	\end{table}

	\begin{table}[]
		\caption{Dissociation temperature ($T_D$) for
			$bb\bar b \bar b$.}\label{tab:t2}
		\centering
		\renewcommand{\arraystretch}{1.4}
		\begin{tabular}{ccccccc}
			\toprule
			& \multicolumn{3}{c}{$\theta=0$} & \multicolumn{3}{c}{$\theta=90$} \\
			\cmidrule(lr){2-4}\cmidrule(lr){5-7}
			$\alpha$ & $\xi=0$      & $\xi=0.1$      & $\xi=0.3$         & $\xi=0$       & $\xi=0.1$       & $\xi=0.3$      \\ 
			\midrule
			0.01                      & 0.4038 $T_c$ & 0.4088 $T_c$ & 0.4193 $T_c$ & 0.4038 $T_c$ & 0.4067 $T_c$ & 0.4121 $T_c$  \\
			0.02                      & 0.4040 $T_c$ & 0.4090 $T_c$ & 0.4195 $T_c$ & 0.4040 $T_c$ & 0.4069 $T_c$ & 0.4124 $T_c$ \\
			0.03                      & 0.4042 $T_c$ & 0.4092 $T_c$  & 0.4198 $T_c$ & 0.4042 $T_c$ & 0.4071 $T_c$ & 0.4126 $T_c$ \\
			1.00                      & 0.4217 $T_c$ & 0.4269 $T_c$ & 0.4379 $T_c$ & 0.4217 $T_c$ & 0.4247 $T_c$ & 0.4304 $T_c$ \\
			\bottomrule
		\end{tabular}
	\end{table}

	\begin{table}[]
		\caption{Masses (GeV) for
			$cc\bar c \bar c$.}\label{tab:t3}
		\centering
		\renewcommand{\arraystretch}{1.3}
		\begin{tabular}{cccccc}
			\toprule
			& \multicolumn{2}{c}{Present} & \multicolumn{3}{c}{Ref.} \\ 
			\cmidrule(lr){2-3}\cmidrule(lr){4-6}
			State & $\alpha=\beta=1$   & $\alpha=\beta=0.8$ & \cite{r9} & \cite{r30} & \cite{r53} \\
			\midrule
			1\textit{s}    & 6.333   & 6.305   & 6.198   & 5.930   & 6.609   \\
			2\textit{s}    & 6.463   & 6.448   &    ---     & 6.202   &    ---     \\
			3\textit{s}    & 6.594   & 6.573   &    ---     & 6.470   &    ---     \\
			4\textit{s}    & 6.707   & 6.681   &    ---     & 6.750   &    ---     \\
			1\textit{p}    & 6.327   & 6.321   &    ---     & 6.204   & 6.611   \\
			2\textit{p}    & 6.477   & 6.462   &   ---      & 6.480   &  --- \\
			\bottomrule     
		\end{tabular}
	\end{table}

	\begin{table}[]
		\caption{Masses (GeV) for
			$bb \bar b \bar b$.}\label{tab:t4}
		\centering
		\renewcommand{\arraystretch}{1.3}
		\begin{tabular}{ccccc}
			\toprule
			& \multicolumn{2}{c}{Present} & \multicolumn{2}{c}{Ref.} \\ 
			\cmidrule(lr){2-3}\cmidrule(lr){4-5}
			State & $\alpha=\beta=1$   & $\alpha=\beta=0.8$ & \cite{r53} & \cite{r30} \\
			\midrule
			1\textit{s}    & 18.647 & 19.682  & 20.012   & 18.650     \\
			2\textit{s}    & 18.702   & 19.730 &   ---     & 18.920 \\
			3\textit{s}    & 18.756 & 19.778 & --- & 19.180 \\
			4\textit{s}    & 18.810 & 19.825 & --- & 19.450  \\
			1\textit{p}    & 18.648 & 19.683 & 20.016 & 18.930  \\
			2\textit{p}    & 18.703 & 19.731 & --- & 19.190 \\
			\bottomrule     
		\end{tabular}
	\end{table}
	
	\begin{table}[]
		\caption{Masses (GeV) for
			$c\bar q c \bar q$.}\label{tab:t5}
		\centering
		\renewcommand{\arraystretch}{1.3}
		\begin{tabular}{ccccccc}
			\toprule
			& \multicolumn{2}{c}{Present} & \multicolumn{4}{c}{Ref.} \\ 
			\cmidrule(lr){2-3}\cmidrule(lr){4-7}
			State & $\alpha=\beta=1$   & $\alpha=\beta=0.1$ & NR\cite{r19}  & \cite{r16} & R\cite{r19} & \cite{r30} \\
			\midrule
			1\textit{s}    & 3.938 & 3.950 & 3.792 & 3.842 & 3.739 & 3.908 \\
			2\textit{s}    & 4.161 & 4.167 & 4.419 & 4.409 & 4.357 & 4.240 \\
			3\textit{s}    & 4.300 & 4.307 & 4.843 & 4.793 & 4.757 & 4.375 \\
			4\textit{s}    & 4.375 & 4.388 & --- & --- & --- & 4.610 \\
			1\textit{p}    & 4.019 & 4.029 & 4.262 & 4.206 & 4.231 & 4.140 \\
			2\textit{p}    & 4.227 & 4.232 & 4.697 & 4.631 & 4.644 & 4.370 \\
			\bottomrule     
		\end{tabular}
	\end{table}

	\begin{table}[]
		\caption{Masses (GeV) for
			$b\bar q b \bar q$.}\label{tab:t6}
		\centering
		\renewcommand{\arraystretch}{1.3}
		\begin{tabular}{ccccccc}
			\toprule
			& \multicolumn{2}{c}{Present} & \multicolumn{4}{c}{Ref.} \\ 
			\cmidrule(lr){2-3}\cmidrule(lr){4-7}
			State & $\alpha=\beta=1$   & $\alpha=\beta=0.4$ & NR\cite{r19}  & \cite{r16} & R\cite{r19} & \cite{r30} \\
			\midrule
			1\textit{s}    & 10.593 & 10.584 & 10.426 & 10.415 & 10.410 & 10.525 \\
			2\textit{s}    & 10.744 & 10.719 & 10.914 & 10.918 & 10.899 & 10.640 \\
			3\textit{s}    & 10.886 & 10.847 & 11.230 & 11.235 & 11.211 & 10.750 \\
			4\textit{s}    & 11.019 & 10.968 & --- & --- & --- & 10.860 \\
			1\textit{p}    & 10.599 & 10.590 & 10.813 & 10.775 & 10.806 & 10.636 \\
			2\textit{p}    & 10.750 & 10.725 & 11.140 & 11.119 & 11.128 & 10.723 \\
			\bottomrule     
		\end{tabular}
	\end{table}

	\clearpage
	\section{Conclusion}
This work provides the solution to the radial SE under a potential explicitly dependent on temperature and the anisotropy parameter. Two cases have studied using GFDNU method: the classical case with ($\alpha = \beta = 1$) and fractional  paradigm ($\alpha,  \beta \neq 1$). 
The dissociation energy of heavy tetraquarks $bb\bar{b}\bar{b}$ and $cc\bar{c}\bar{c}$ in the 1\textit{p} state has computed in the fractional case and then we get the classical case as a particular scenario at $\alpha = \beta = 1$. We have plotted the dissociation energy with temperature under the effect of the anisotropic parameter in two cases: $\theta = 0$ and $\theta = \pi/2$.
It was discovered that when temperature increased, dissociation energy reduced but anisotropic parameter increased. Also, the dissociation energy was smaller in fractional  paradigm than in the conventional paradigm. 
In the realm of fractional calculus and quantum mechanics, the fractal features of space-time are represented by the parameter $\alpha$ and in this context it plays a significant role with regard to the solutions of the SE. The physical effect of this parameter on the character of wave functions. Although it is not a physical parameter, the value of $\alpha$ can be changed to allow researchers to understand the differences between quantum and classical systems in fractal environments. 
In addition, we studied $T_D$ of $cc\bar{c}\bar{c}$ and $bb\bar{b}\bar{b}$ in classical and fractional case under effect the fraction parameter and anisotropic parameter in two case $\theta=0$ and $\theta=\frac{\pi}{2}$ and we noted that dissociation energy is lower at $\theta=90$ compared with $\theta=0$.  
	Also, by increasing fraction parameter dissociation energy increased and dissociation energy increased in anisotropic parameter than isotropic parameter.
	Finally, we calculated the mass of $cc\bar{c}\bar{c}$, $bb\bar{b}\bar{b}$, $c\bar{q}c\bar{q}$ and $b\bar{q}b\bar{q}$ in two cases: the fraction case and classical case. and we compare this results with previous studies and this results are agreement of with Refs. \cite{r9}, \cite{r16}, \cite{r19}, \cite{r30}, \cite{r53}.
Therefore, by analyzing the combined impacts of fractional space-time dynamics and hot plasma anisotropy, this work bridges a long-standing gap in exotic multiquark literature, providing rigorous theoretical constraints to guide future experimental searches of exotic matter production at global facilities such as the LHC and RHIC.
\\
\\
{\large\textbf{ Data availability }}\\
All data generated or analyzed during this study are available upon reasonable request from the corresponding author.
\\
\\
{\large\textbf{ Conflict of interest }}\\
The authors declare that they have no competing interests.
\\
\\
{\large\textbf{ Author contributions }}\\
All authors contributed to the work’s conception and design. All authors read and approved the final manuscript.
\\
\\
{\large\textbf{Funding}}\\
No funding was received for conducting this study.


\clearpage	
	\begin{thebibliography}{99}
		\bibitem{r1} S.-K. Choi et al. (Belle Collaboration),
		``Observation of a narrow charmoniumlike state in exclusive $B^{\pm} \to K^{\pm}\pi^{+}\pi^{-}J/\psi$ decays,''
		Physical Review Letters, vol. 91, no. 26, p. 262001, 2003.
		
		\bibitem{r2} M. Ablikim et al. (BESIII Collaboration),
		``Observation of a charged charmoniumlike structure in $e^{+}e^{-} \to \pi^{+}\pi^{-}J/\psi$ at $\sqrt{s} = 4.26~\text{GeV}$,''
		Physical Review Letters, vol. 110, no. 25, p. 252001, 2013.
		
		\bibitem{r3} S.-K. Choi et al. (Belle Collaboration),
		``Observation of a resonance-like structure in the $\pi^{\pm}\psi'$ mass distribution in exclusive $B \to K\pi^{\pm}\psi'$ decays,''
		Physical Review Letters, vol. 100, no. 14, p. 142001, 2008.		
		
		\bibitem{r4} The CMS Collaboration, "Determination of the spin and parity of all-charm tetraquarks," Nature, vol. 648, pp. 58–63, 2025.
		
		\bibitem{r5}  S. S. Agaev, K. Azizi, B. Barsbay, and H. Sundu, "Exploring fully heavy scalar tetraquarks $QQ\bar Q\bar Q$," Physics Letters B, vol. 844, p. 138089, 2023.
		
		\bibitem{r6} Z. Ghalenovi, "Masses of charmed and bottom tetraquarks in the non-relativistic quark model," International Journal of Modern Physics: Conference Series, vol. 46, p. 1860084, 2018.
		
		\bibitem{r7} V. R. Debastiani and F. S. Navarra, "A non-relativistic model for the [$cc$][$\bar  c \bar c$] tetraquark," Chin. Phys. C, vol. 43, p. 013105, 2019.
		
		\bibitem{r8}  R. Mistry and A. Majethiya, "Spectroscopic study of exotic fully-heavy tetraquark states $QQ\bar Q\bar Q$ $(Q\in{c, b})$," Chinese Journal of Physics, vol. 91, pp. 932–946, 2024.
		
		\bibitem{r9}  P. Lundhammar and T. Ohlsson, "Nonrelativistic model of tetraquarks and predictions for their masses from fits to charmed and bottom meson data," Physical Review D, vol. 102, no. 5, p. 054018, 2020.
		
		\bibitem{r10}  R. Tiwari, J. Oudichhya, and A. K. Rai, "Mass-spectra of light-heavy tetraquarks," International Journal of Modern Physics A, vol. 38, no. 33n34, p. 2341007, 2023.
		
		\bibitem{r11}  T. Bhavsar, M. Shah, S. Patel, and P. C. Vinodkumar, "Masses of tetraquark states in the hidden charm sector above $D$-$D^*$ threshold," Nuclear Physics A, vol. 1000, p. 121856, 2020.
		
		\bibitem{r12} R. Jamshidzadeh, M. Monemzadeh, and N. Tazimi, "Calculating the binding energies and masses of fully heavy tetraquarks," Physica Scripta, vol. 100, no. 7, p. 075304, 2025. 
		
		\bibitem{r13} H. Mutuk, "Spectrum of $cc\bar b\bar b$, $bc\bar c\bar c$, and $bc\bar b\bar b$ tetraquark states in the dynamical diquark model," Physics Letters B, vol. 834, p. 137404, 2022.
		
		\bibitem{r14} Q. Meng, G. J. Wang, and M. Oka, "Mass spectra of full-heavy and double-heavy tetraquark states in the conventional quark model," Physical Review D, vol. 111, no. 1, p. 014018, 2025.
		
		\bibitem{r15} A. Ghasempour, N. Tazimi, and M. Monemzadeh, "Analytical investigation of the mass spectrum of tetraquarks with a relativistic approach," The European Physical Journal C, vol. 85, no. 1, p. 113, 2025.
		
		\bibitem{r16} F. Zolfagharpour and M. Aslanzadeh, "Heavy tetraquarks in a relativistic diquarkonium picture," The European Physical Journal A, vol. 55, no. 6, p. 86, 2019.
		
		\bibitem{r17} N. Shiri, N. Tazimi, and M. Monemzadeh, "Analytical solution of relativistic four quark bound systems," The European Physical Journal C, vol. 83, no. 1, p. 53, 2023.
		
		\bibitem{r18} R. N. Faustov, V. O. Galkin, and E. M. Savchenko, "Masses of the $QQ\bar Q\bar Q$ tetraquarks in the relativistic diquark–antidiquark picture," Physical Review D, vol. 102, no. 11, p. 114030, 2020.
		
		\bibitem{r19} M. R. Hadizadeh and A. Khaledi-Nasab, "Heavy tetraquarks in the diquark–antidiquark picture," Physics Letters B, vol. 753, pp. 8–12, 2016.
		
		\bibitem{r20} V. Patel, J. Oudichhya, and A. K. Rai, "Spectroscopic study of $ss\bar b\bar b$ and $bb\bar b\bar b$ tetraquarks using Regge phenomenology," International Journal of Modern Physics A, vol. 40, no. 31, p. 2550152, 2025.
		
		\bibitem{r21} V. Patel, J. Oudichhya, and A. K. Rai, "Spectroscopy of $cc\bar c\bar c$ and $ss\bar c\bar c$ tetraquarks within the framework of Regge phenomenology," Few-Body Systems, vol. 66, no. 4, p. 39, 2025.
		
		\bibitem{r22} M. N. Anwar and T. J. Burns, "Tetraquark mass relations in quark and diquark models," Physics Letters B, vol. 847, p. 138248, 2023.
		
		\bibitem{r23} A. Ghasempour, N. Tazimi, and M. Monemzadeh, "Analytical investigation of the mass spectrum of tetraquarks with a relativistic approach," The European Physical Journal C, vol. 85, no. 1, p. 113, 2025.
		
		\bibitem{r24} M. Strickland and D. Bazow, "Thermal bottomonium suppression at RHIC and LHC," Nuclear Physics A, vol. 879, pp. 25–58, 2012.
		
		\bibitem{r25} L. Thakur, N. Haque, U. Kakade, and B. K. Patra, "Dissociation of quarkonium in an anisotropic hot QCD medium," Physical Review D, vol. 88, p. 012007, 2013.
		
		\bibitem{r26} J. Zhao, S. Shi, and P. Zhuang, "Fully-heavy tetraquarks in a strongly interacting medium," Physical Review D, vol. 102, no. 11, p. 114001, 2020.
		
		\bibitem{r27} A. Aydın, H. Sundu, J. Y. Süngü, and E. V. Veliev, "Scalar fully-charm and bottom tetraquarks under extreme temperatures," The European Physical Journal C, vol. 85, no. 5, p. 567, 2025.
		
		\bibitem{r28} V. S. Silva, C. Pigozzo, and L. M. Abreu, "Fully-heavy tetraquarks in the vacuum and in a hot environment," The European Physical Journal C, vol. 85, no. 7, p. 783, 2025.
		
		\bibitem{r29} M. Abu-Shady and H. M. Fath-Allah, "Masses of single, double, and triple heavy baryons in the hyper-central quark model by using GF-AEIM," Advances in High Energy Physics, vol. 2022, Art. no. 4539308, 2022.
		
		\bibitem{r30} M. Abu-Shady, M. M. A. Ahmed, and N. H. Gerish, "Generalized fractional of the extended Nikiforov–Uvarov method for heavy tetraquark masses spectra," Modern Physics Letters A, vol. 38, no. 4, p. 2350028, 2023.
		
		\bibitem{r31} M. Abu-Shady and S. Y. Ezz-Alarab, "Thermodynamic properties of heavy mesons in strongly coupled quark gluon plasma using the fractional of non-relativistic quark model," Indian Journal of Physics, vol. 97, no. 12, pp. 3661–3677, 2023.
		
		\bibitem{r32} M. Abu-Shady, "Studying quarkonium in the anisotropic hot-dense quark-gluon plasma medium in the framework of generalized fractional derivative," Revista mexicana de física, vol. 69, no. 4, Art. no. 040701, 2023. 
		
		\bibitem{r33} E. Omugbe, M. Abu-Shady, and E. P. Inyang, "Approximate bound state solutions of the fractional SE under the spin-spin-dependent Cornell potential," Journal of the Nigerian Society of Physical Sciences, vol. 6, p. 1771, 2024. 
		
		\bibitem{r34} M. Abu-Shady and H. M. Fath-Allah, "Properties and behaviors of heavy quarkonia: insights through fractional  paradigm and topological defects," Advances in High Energy Physics, vol. 2024, Art. no. 2730568, 2024.
		
		\bibitem{r35} M. Abu-Shady, H. Ahmad, H. Alotaibi, and A. R. Ali, "Investigating the fractional wave function and the impact of topological defects with anisotropic plasma on the dissociation of bottomonium in the fractional non-relativistic quark model," AIP Advances, vol. 14, Art. no. 045011, 2024.
		
		\bibitem{r36} M. Abu-Shady and H. M. Fath-Allah, "Investigating heavy quarkonia binding in an anisotropic-dense quark-gluon plasma with topological defects in the framework of fractional non-relativistic quark model," Scientific Reports, vol. 15, no. 1, p. 1875, 2025.
		
		\bibitem{r37} D. N. Ongodo, A. A. Likéné, J. E. A. Ema'a, P. E. Abiama, and G. H. Ben-Bolie, "Hyperfine mass splittings in ground and radially excited states of heavy-flavored  $QQ\bar Q \bar Q$ tetraquarks: PGM defect and fractional order effects," Nuclear Physics A, vol. 1063, p. 123215, 2025.
		
		\bibitem{r38} D. N. Ongodo, A. A. Likéné, J. M. Ema’a, P. Abiama, and G. H. Ben-Bolie, "Effect of spin-spin interaction and fractional order on heavy pentaquark masses under topological defect space-times," The European Physical Journal C, vol. 85, no. 4, pp. 1–13, 2025. 
		
		\bibitem{r39}	M. Abu-Shady and M. K. Kaabar, “A generalized definition of the fractional derivative with applications,” Mathematical Problems in Engineering, vol. 2021, no. 1, Art. no. 9444803, 2021.
		
		\bibitem{r40} M. M. Hammad, A. S. Yaqut, M. A. Abdel-Khalek, and S. B. Doma, "Analytical study of conformable fractional Bohr Hamiltonian with Kratzer potential," Nuclear Physics A, vol. 1015, Art. no. 122307, 2021.
		
		\bibitem{r41} S. M. Kuchin and N. V. Maksimenko, "Theoretical estimations of the spin-averaged mass spectra of heavy quarkonia and Bc mesons," Universal Journal of Physics and Application, vol. 7, no. 3, pp. 295-298, 2013.
		
		\bibitem{r42} A. F. Nikiforov and V. B. Uvarov, Special Functions of Mathematical Physics, vol. 205. Basel, Switzerland: Birkhäuser, 1988.
		
		\bibitem{r43} M. Abu-Shady and H. M. Fath-Allah, "The parametric generalized fractional Nikiforov-Uvarov method and its applications," East European Journal of Physics, no. 3, pp. 248–262, 2023.
		
		\bibitem{r44} C. Tezcan and R. Sever, "A general approach for the exact solution of the SE," International Journal of Theoretical Physics, vol. 48, no. 2, pp. 337–350, 2009.
		
		\bibitem{r45} P. Petreczky, "Quarkonium in a hot medium," Journal of Physics G: Nuclear Physics, vol. 37, no. 9, Art. no. 094009, 2010.
		
		\bibitem{r46} V. K. Agotiya, V. Chandra, M. Y. Jamal, and I. Nilima, "Dissociation of heavy quarkonium in hot QCD medium in a quasiparticle model," Physical Review D, vol. 94, no. 9, Art. no. 094006, 2016.
		
		\bibitem{r47} V. Kumar, "Spectrum Analysis of Mesons using Nikiforov-Uvarov Functional Analysis Method," Materials Research Proceedings, vol. 22, pp. 7–12, Mar. 2022, doi: 10.21741/9781644901878-2.
		
		\bibitem{r48} F. Karsch, M. T. Mehr, and H. Satz, "Color screening and deconfinement for bound states of heavy quarks," Zeitschrift für Physik C: Particles and Fields, vol. 37, no. 4, pp. 617–622, 1988.
		
		\bibitem{r49} H. Satz, "Charm and beauty in a hot environment," 2006, arXiv:hep-ph/0602245.
		
		\bibitem{r50} T. Song, K. C. Han, and C. M. Ko, "Charmonium production in relativistic heavy-ion collisions," Physical Review C—Nuclear Physics, vol. 84, no. 3, Art. no. 034907, 2011.
		
		\bibitem{r51} D. Lafferty and A. Rothkopf, "Improved Gauss law model and in-medium heavy quarkonium at finite density and velocity," Physical Review D, vol. 101, no. 5, Art. no. 056010, 2020.
		
		\bibitem{r52} M. Abu-Shady, H. M. Mansour, and A. I. Ahmadov, "Dissociation of quarkonium in hot and dense media in an anisotropic plasma in the nonrelativistic quark model," Advances in High Energy Physics, vol. 2019, no. 1, Art. no. 4785615, 2019.		
		
		\bibitem{r53} M. Abu-Shady, M. M. A. Ahmed, and N. H. Gerish, "The non-relativistic treatment of heavy tetraquark masses in the logarithmic quark potential," Revista Mexicana de Física, vol. 68, no. 6, Art. no. 061201, 2022.
		
		
		
		
		
		
		
	\end{thebibliography}
\end{document}